\documentclass[usegraphicx,useAMS]{mn2e}

\usepackage{amssymb}  % for \geqslant

\newcommand\teff{\mbox{$T_{\rm eff}$}}

\def\eps@scaling{1.0}%
\newcommand\epsscale[1]{\gdef\eps@scaling{#1}}%

\newcommand\plotone[1]{%
 \centering
 \leavevmode
 \includegraphics[width={\eps@scaling\columnwidth}]{#1}%
}%

\newcommand\plottwo[2]{%
 \centering
 \leavevmode
 \columnwidth=.48\textwidth
 \includegraphics[width={\eps@scaling\columnwidth}]{#1}%
 \hfil
 \includegraphics[width={\eps@scaling\columnwidth}]{#2}%
}%

\title[Synthetic $\Delta a$ photometry]{The 5200~{\AA} flux depression
       of chemically peculiar stars: I.~Synthetic $\Delta a$ photometry - the normality line.}
\author[Kupka et al.]
       {F.~Kupka,$^{1,2}$ E.~Paunzen,$^1$\thanks{Affiliated to
           Zentraler Informatikdienst der Universit\"at Wien,
           Universit\"atsstr. 7, A-1010 Wien,
           e-mail: Ernst.Paunzen@univie.ac.at} and
       H.M.~Maitzen$^1$ \\ 
       $^1$Institut f\"ur Astronomie, Universit\"at Wien,
       T\"urkenschanzstra{\ss}e 17, A-1180 Wien, Austria
       (f.kupka@qmul.ac.uk, maitzen@astro.univie.ac.at) \\
       $^2$Astronomy Unit, School of Mathematical Sciences, 
       Queen Mary, University of London, Mile End Road,
       London E1 4NS, UK 
}

\date{Submitted 2002 October 23.}

\pagerange{\pageref{firstpage}--\pageref{lastpage}}
\pubyear{2003}

\begin{document}

\maketitle

\label{firstpage}

\begin{abstract}

The $\Delta a$ photometric system provides an efficient observational method to 
identify and distinguish magnetic and several other types of 
chemically peculiar (CP) stars of spectral types B to F from other 
classes of stars in the same range of effective temperatures. We have 
developed a synthetic photometric system which can be used to explore the 
capability of model atmospheres with individual element abundances to 
predict photometric $\Delta a$ magnitudes which measure the extent of the 
flux depression around 5200~{\AA} found in different types of CP stars. 
In this first paper, we confirm the observed dependency of the $a$ index as a 
function of various colour indices sensitive to the effective temperature of stars 
as well as its average scatter expected from surface gravity variations within the 
main sequence band. The behaviour of the so-called ``normality line'' of $\Delta a$ 
systems used in photometric observations of CP stars is well reproduced. The metallicity 
dependence of the normality line of the $\Delta a$ system was computed for several 
grids of model atmospheres where the abundances of elements heavier than He had been scaled 
$\pm$0.5\,dex with respect to the solar value. We estimate a lowering of 
$\Delta a$ magnitudes for CP 
stars within the Magellanic Clouds by $\sim -3$~mmag relative to those in the solar 
neighbourhood assuming an average metallicity of $[{\rm Fe}/{\rm H}]= -0.5$~dex. Using 
these results on the metallicity bias of the $\Delta a$ system we find the observational 
systems in use suitable to identify CP 
stars in other galaxies or distant regions of our own galaxy and capable to provide 
data samples on a statistically meaningful basis. In turn, the synthetic system is 
suitable to test the performance of model atmospheres for CP stars. This work will be 
presented in follow-up papers of this series.

\end{abstract}


\section{Introduction}   \label{Sect1}

The chemically peculiar (CP) stars of the upper main sequence have
been targets for astrophysical studies since the discovery of these
objects by the American astronomer Antonia Maury (1897). Most of this
early research was devoted to the detection of peculiar features in
their spectra and photometric behaviour. The main characteristics of
the classical CP stars are: peculiar and often variable line strengths,
quadrature of line variability with radial velocity changes, photometric
variability with the same periodicity and coincidence of extrema. Slow
rotation was inferred from the sharpness of spectral lines.
Overabundances of several orders of magnitude compared to the Sun were
derived for Si, Cr, Sr, Eu, and for other heavy elements.

Babcock (1947) discovered a global dipolar magnetic field in the star
78~Virginis followed by a catalogue of similar stars (Babcock 1958) in which
also the variability of the field strength in many CP stars --- including
even a reversal of magnetic polarity --- was discovered. Stibbs (1950)
introduced the Oblique Rotator concept of slowly rotating stars with
non-coincidence of the magnetic and rotational axes. This model
reproduces variability and reversals of the magnetic field strength.
Due to the chemical abundance concentrations at the magnetic
poles also spectral and the related photometric variabilities are
easily understood, as well as radial velocity variations of the appearing
and receding patches on the stellar surface.

Preston (1974) divided the CP stars into the following groups:
\begin{itemize}
\item CP1: Am/Fm stars without a strong global magnetic field; weak lines
of Ca\,{\sc ii} and Sc\,{\sc ii}, otherwise strong overabundances;
\item CP2: ``classical'' CP stars with strong magnetic fields (they are
      also known as the magnetic CP or mCP stars);
\item CP3: HgMn stars, basically non-magnetic;
\item CP4: He-weak stars, some of these objects show a detectable
           magnetic field.
\end{itemize}

Kodaira (1969) was the first who noticed significant flux depressions at
4100\,\AA, 5200\,\AA, and 6300\,\AA\, in the spectrum of HD~221568. Jamar
(1977, 1978) investigated similar features in the ultraviolet region
at 1400\,\AA, 1750\,\AA, and 2750\,\AA. All features were found to be only
visible in magnetic CP stars. Maitzen (1976) introduced the narrow
band, three filter $\Delta a$ system in order to investigate the flux
depression at 5200\,\AA. It samples the depth of this flux depression
by comparing the flux at the center (5220~{\AA}, $g_{\rm 2}$),
with the adjacent regions (5000~{\AA}, $g_{\rm 1}$ and 5500~{\AA},
$y$) using a band-width of 130~{\AA} (for $g_{\rm 1}$ and $g_{\rm 2}$)
and the band-width of 230~{\AA} for the Str\"omgren y filter. The respective
index was introduced as: $$ a = g_{\rm 2} - (g_{\rm 1} + y)/2 $$
Since this quantity is slightly dependent on temperature (increasing
towards lower temperatures), the intrinsic peculiarity index
had to be defined as
$$ \Delta a = a - a_{\rm 0}[(b - y); (B - V); (g_{\rm 1} - y)] $$
i.e. the difference between the individual $a$-value and the $a$-value
of non-peculiar stars of the same colour. The locus of the $a_{\rm 0}$-values
for non-peculiar objects has been called normality line. It was shown 
(e.g. Vogt et al. 1998) that
virtually all peculiar stars with magnetic fields (CP2 stars) have positive
$\Delta a$ values up to $+75$\,mmag whereas Be/Ae and $\lambda$ Bootis stars
exhibit significantly negative ones (Maitzen \& Pavlovski 1989a,b,c). Extreme
cases of the CP1 and CP3 group may exhibit marginally peculiar positive
$\Delta a$ values (Maitzen \& Vogt 1983).

Several attempts have been made to explain the origin of this feature.
Adelman \& Wolken (1976) and Adelman, Shore \& Wolken (1976) investigated bound-free
discontinuities, Jamar, Macau-Hercot \& Praderie (1978) proposed autoionisation transitions
of Si\,{\sc ii}, whereas enhanced line absorption was discussed by Maitzen
(1976) and Maitzen \& Muthsam (1980). The latter presented a comparison
of synthesised flux distributions and observed spectrophotometry (the first
attempt in this direction was by Leckrone, Fowler \& Adelman 1974). From their
synthetic spectra they recovered a narrow and deep feature at
about 5175\,\AA\, and a broad component centred at about 5275\,\AA. They
were not able to reproduce the flux depression for effective temperatures
higher than 8000\,K. More recently, Adelman et al.\ (1995) have used model
atmospheres of Kurucz (1993a,b) with enhanced metallicity (i.e.\ a solar
element abundance distribution where all elements heavier than He had been
scaled by +0.5 or +1.0 dex) and concluded that at least part of the
5200~{\AA} feature in magnetic CP stars may be due to differential line
blanketing. In a follow-up work, Adelman \& Rayle (2000) have extended this
study to a larger group of stars and found that solar composition models may
successfully predict the flux distribution of normal and many CP3 (HgMn)
stars, while they fail to do so for a number of CP2 stars, i.e.\ the group of
stars showing the largest flux depression in the 5200~{\AA} region and thus
the largest magnitudes in $\Delta a$.

One of the main conclusions drawn from these previous works has been
the necessity to build specific model stellar atmospheres for CP stars
using state-of-the-art opacity data. The increase in available computer
power and advances in computational algorithms during the last two decades
have now brought this problem into the realm of workstations and personal
computers. Moreover, stellar atmosphere modelling can take advantage of
data bases for atomic line transitions devoted to stellar atmosphere
applications such as Kurucz (1992) and the VALD project (Kupka et al.\ 1999;
Ryabchikova et al.\ 1999). Among the current projects for the computation
of model atmospheres there are several which can calculate models with
individual abundances on workstations or personal computers. Two of them are
based on an opacity sampling approach. ATLAS12 by Kurucz (1996), for which
a first application was presented by Castelli \& Kurucz (1994), is
particularly suitable for B to K type stars at or close to the main sequence.
The marcs project in Uppsala (Bengt Gustafsson and his group) is aimed at
the cool part of the Hertzsprung-Russell diagram and can produce model atmospheres for stars
with spectral types later than A0 (Gustafsson et al. 2003). 
However, for the computation of small grids of model
atmospheres with individual abundances, which are required when varying
\teff\ or $\log(g)$ during the initial analysis of a single star or several
sufficiently similar stars, the opacity distribution function
(ODF) approach remains preferable due to its higher computational
efficiency. Piskunov \& Kupka (2001) have presented a new software toolkit
based on this approach using a modified version of the ATLAS9 code of
Kurucz (1993b). It is suitable for spectral types from early B type to
early F type stars at or close to the main sequence. 

Our study has been initiated to synthesise and reproduce the 
characteristics of the
5200\,\AA\, flux depression which is measured via
the $\Delta a$ photometric system with the
currently available stellar atmosphere models.
Our general aim is to explain the various aspects
and characteristics of the $\Delta a$ photometric system
(and thus the 5200\,\AA\, flux depression),
i.e. the dependency on the metallicity, surface
gravity and effective temperature. 

The new synthetic photometric $\Delta a$ system is introduced in Sect.~\ref{Sect2}.
We discuss
the chosen filter system and its properties with respect to systems that have
been used for observations. We compare the $\Delta a$ calibration relations
obtained from grids of standard model atmospheres with observed relations as
a function of several photometric indicators of \teff. We also discuss
luminosity and metallicity effects and compute the expected zero point shift
of the normality line $a_0$ of $\Delta a$ photometry when applying the latter
to the Magellanic Clouds. Following the standard used in the literature we
give values for $\Delta a$ in units of mmag throughout this paper, while
other photometric quantities are given in mag with normalisations as
described in Sect.~\ref{Sect2}. Whenever the unit of mmag is used, it
is explicitly mentioned so in the text to avoid confusion. In the concluding
Sect.~\ref{Sect3} we summarise the success of the synthetic $\Delta a$ system
in reproducing the normality line and the usefulness of $\Delta a$ photometry
to study CP stars in environments different from the solar neighbourhood. We
also provide an outlook for the next papers of this series in which
the software and methods of Piskunov \& Kupka (2001) will be used. 
In two follow-up papers we will discuss the
capability and limitations of homogeneous model atmospheres with individual abundances for
reproducing observed $\Delta a$ indices of different types of CP stars.

\section{Synthetic photometry}   \label{Sect2}

\subsection{The synthetic $\Delta a$ filter system}

Relyea \& Kurucz (1978) have discussed in detail how to calculate synthetic
colours from model atmosphere fluxes computed with the ATLAS code. The
main idea is to convolve the emergent surface fluxes predicted from model
atmospheres with several functions representing filter transmission,
relative absorption of all other devices in the optical path (including
telescope mirrors), and detector sensitivity.

\begin{figure}
\plotone{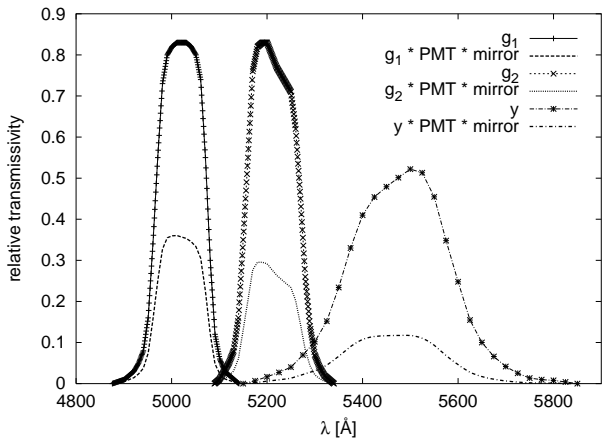}
\caption{Filter transmission functions and effective transmission of
         the synthetic photometric system after convolution with the profiles
         of the response function of a 1P21 RCA photomultiplier tube (PMT)
         and a typical mirror reflection efficiency function (the latter
         two are taken from the {\sc uvby.for} programme of Kurucz 1993b).}
\label{Fig.filters}
\end{figure}

\begin{figure*}
\plottwo{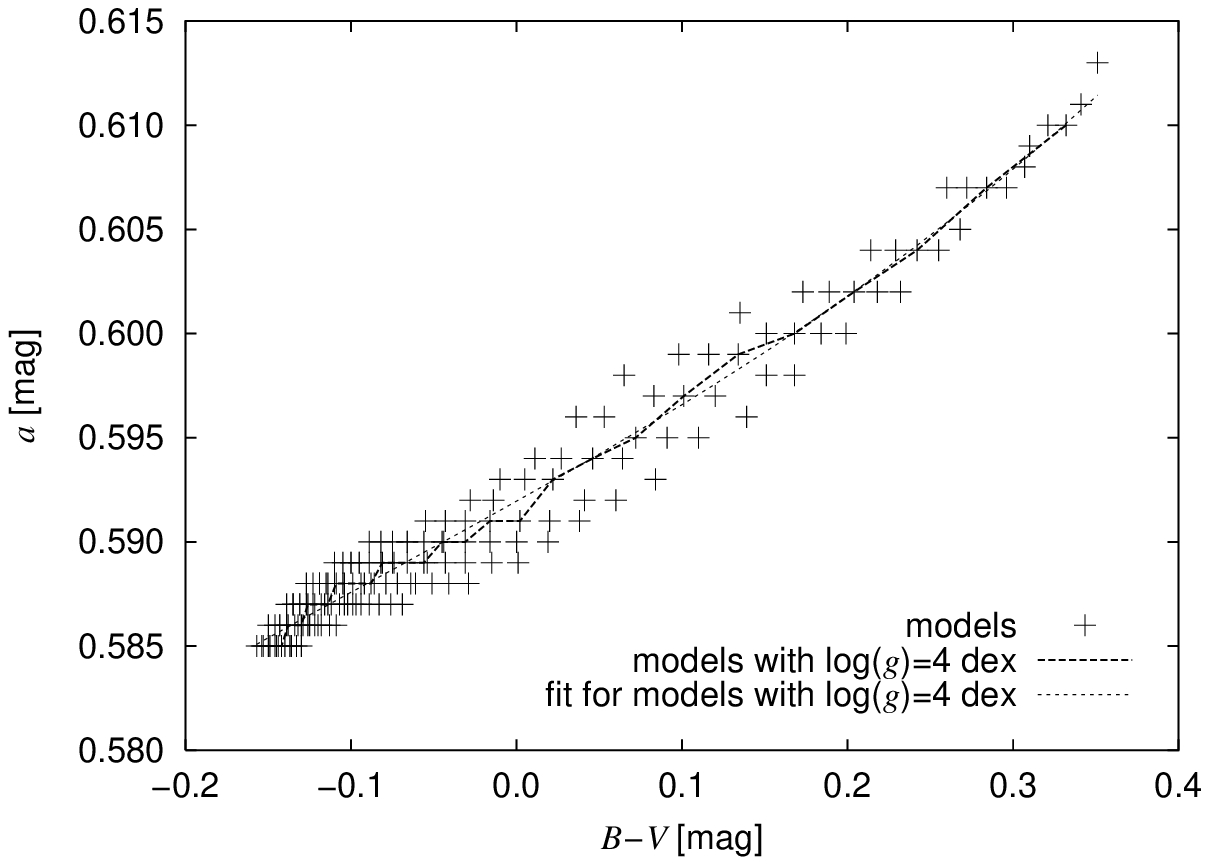}{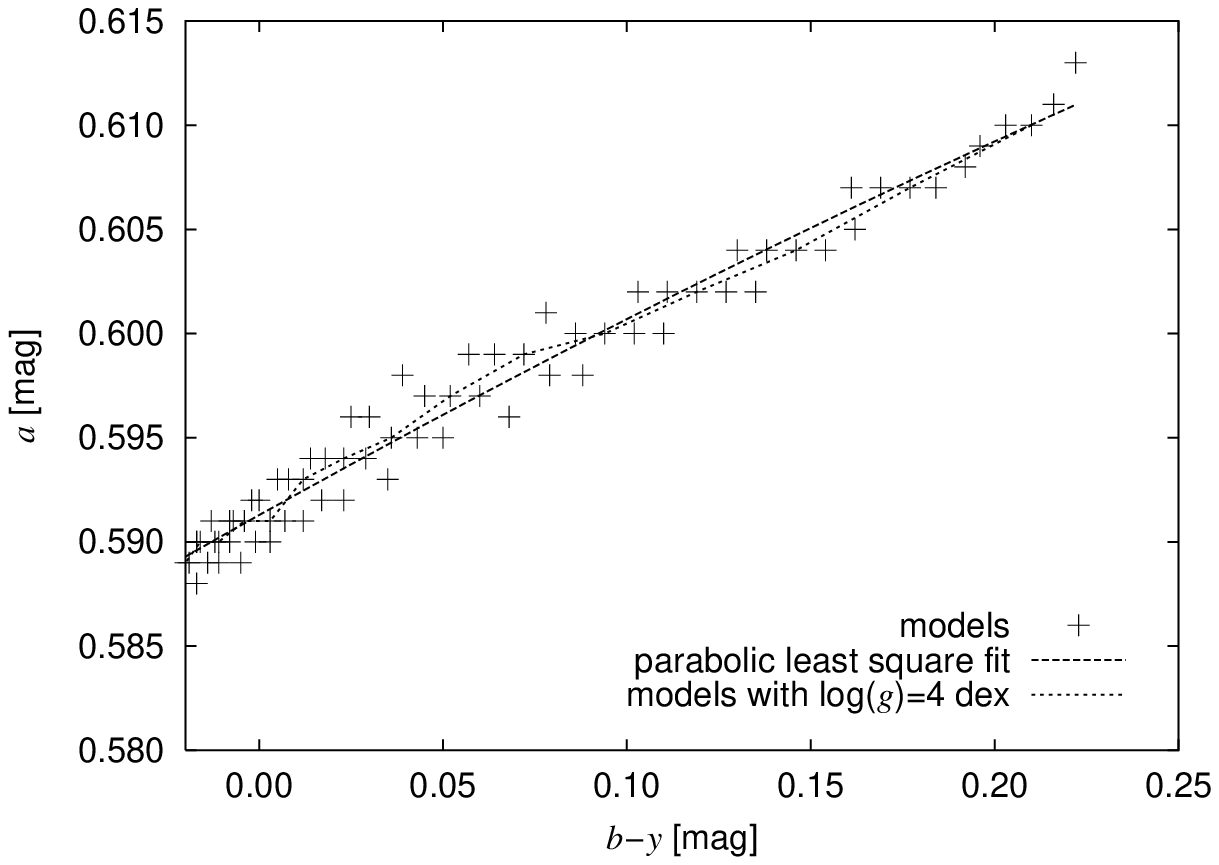}
\plottwo{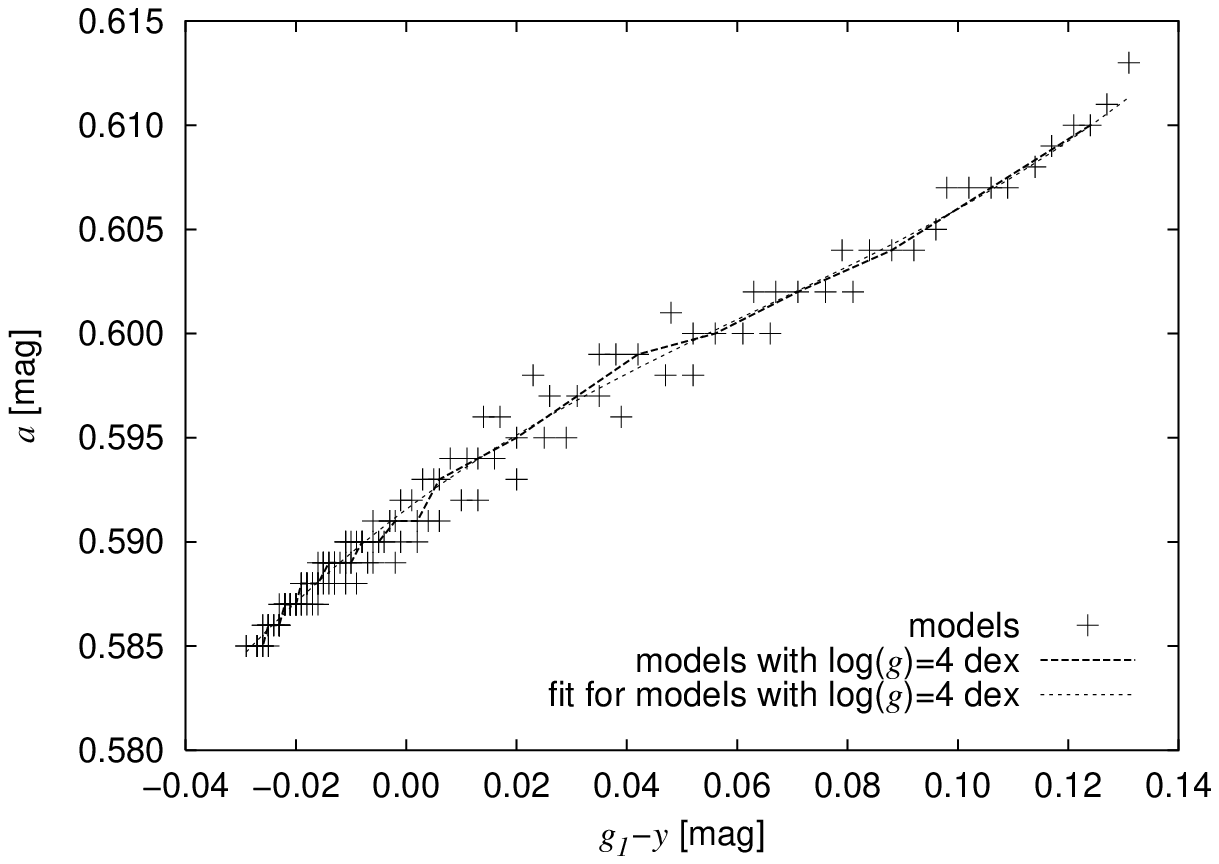}{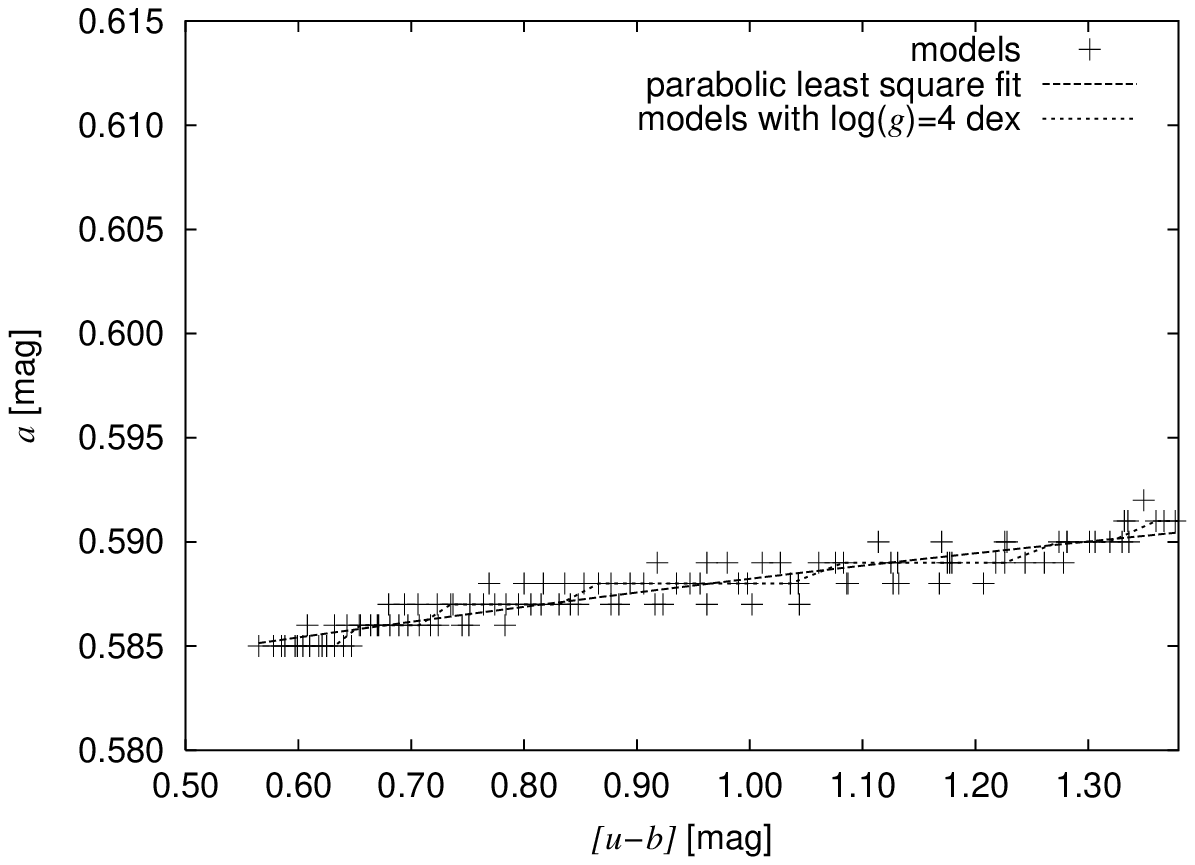}
\caption{Temperature and gravity dependence along the main sequence for
         various observational indicators of \teff. Colours were computed
         from ATLAS9 type model atmospheres with \teff=[7000,15000]~K for
         $\log(g)$\,=\,[3.5,4.5]~dex and solar metallicity. For the upper right
         panel, models were restricted to $(b-y)\geq-0.02$ while for the
         lower right panel, models were restricted to $[$u-b$]\leq 1.38$. The
         zero point of the $(g_1-y)$ colour was set to coincide with the
         zero point of $(b-y)$ at a $\log(g)$ of 4~dex (a constant of
         0.269 had to be added to the output value of $(g_1-y)$). Models
         with $\log(g)$ of 4~dex have been connected by straight lines
         to indicate the shape, zero-point, and slope of the normality
         line. For the cooler models (towards the right-hand side within
         each panel) the true ZAMS and thus the normality line $a_0$ are
         located slightly below this line (cf.\ Fig.~\ref{Fig.lumin_calib}).
         For the left-side panels a cubic least square fit through the
         $\log(g)=4$~dex models is displayed, while right-side ones include
         a parabolic fit through all the models displayed.}
\label{Fig.colour_calib}
\end{figure*}

Figure~\ref{Fig.filters} shows the response functions of our synthetic
photometric system. The steep decay of the response function of the
1P21 RCA photomultiplier from 58\% at 5000~{\AA} to 29\% at 5500~{\AA}
and a mere 10\% at 6000~{\AA} explains the smaller effective sensitivity
of the system in the $y$-band relative to the $g_1$-band in comparison
with the transmission functions of the filters themselves. We note
here that the transmission functions for the $g_1$ and $g_2$ filters
were taken from calibration measurements of one of the filter sets used
in earlier observations (System ``2'' in Maitzen \& Vogt 1983), while the 
$y$ filter transmission (difference less than 1\,\% to that of Maitzen
\& Vogt 1983)
and the mirror reflection efficiency and detector response
functions were taken from the {\sc uvby} code of Kurucz (1993b). 
Maitzen \& Vogt (1983) used the response function
of an EMI 6256 photomultiplier for their calibrations which has essentially 
the same shape but a different
absolute sensitivity as the 1P21 one in the relevant wavelength region. This
offset in the absolute sensitivity does not affect our synthetic magnitudes
since we only use differences between individual filters.

We note that all $\Delta a$ measurements used for comparisons in this paper were observed
within the classical photomultiplier system. The new CCD $\Delta a$ system
(Maitzen, Paunzen \& Rode 1997; Bayer et al.\ 2000; Maitzen et al. 2001;
Paunzen \& Maitzen 2001, 2002; Paunzen et al. 2002) 
in turn has been modified in
order to compensate for the different response functions of a photomultiplier
and a CCD. A classical photomultiplier is almost insensitive at approximately
6000\,{\AA} whereas a CCD is very sensitive in the red region. This implies
an almost linear increase of the response function from $g_1$ to $y$.
In the new CCD system, the FWHM of the $g_1$ and $y$ filters are 222\,{\AA}
and 120\,{\AA}, respectively. This guarantees that the total flux of each
filter after the convolution with the response function is comparable with
that one of the ``old'' system.

\subsection{Calibration relations}  \label{Sect_calib}

The overall success of the ATLAS9 model atmospheres of Kurucz (1993b) to
reproduce photometric colours and spectrophotometric fluxes of standard stars
of spectral types B and A (Castelli \& Kurucz 1994; Smalley \& Dworetsky
1995; Castelli, Gratton \& Kurucz 1997; Castelli 1998) and also for some of the CP
stars (Adelman \& Rayle 2000) promotes them as a logical choice when testing
a synthetic photometric system. We have thus computed several grids of
ATLAS9 model atmospheres with the Stellar Model Grid Tool (SMGT; see Heiter et al.\ 2002 
for a description) using the line opacities from Kurucz
(1993a) and the ATLAS9 code of Kurucz (1993b), unaltered except for the
convection treatment (Smalley \& Kupka 1997; Heiter et al.\ 2002). 
In fact, both Smalley \& Kupka (1997) and Heiter et al.\ (2002)
recommend the use of a convection model in ATLAS9 which predicts
inefficient convection for mid to late A stars (\teff\ $\leqslant 8500$~K).
We thus used the model of Canuto \& Mazzitelli (1991) which allows
a better reproduction of Str\"omgren colours of A stars than the original
models of Kurucz (1993b), as shown in Smalley \& Kupka (1997).
For
models with \teff\ $> 8500$~K, where convection has only negligible influence
on temperature gradients and colours, our models are virtually identical
to those from the original grids published by Kurucz (1993b).
Anyway,
from a comparison we did with model atmospheres based on different
convection models we conclude that for any of the convection treatments
available for ATLAS9 (cf.\ Castelli et al.\ 1997; Heiter et al.\ 2002)
the $a$ values change only by 0 to $+3$~mmag for the coolest models in our
grids and remain completely unaltered for models with \teff\ $> 8500$~K.
Thus, no important bias is introduced into our calibration tests by selecting
a particular convection model. The ATLAS9 grids of Kurucz (1993b)
ones have a spacing of $\Delta \log(g)\,=\,0.5$~dex which is slightly too
coarse for our purpose. Hence, we also included intermediate models in our
computations by using a spacing of $\Delta \log(g)\,=\,0.25$~dex.
We studied a $\log(g)$--range from 2.5~dex to 4.5~dex and a \teff--range from
7000~K to 15000~K (with a spacing of 250~K as in Kurucz 1993b). Model
atmospheres assuming one of the following three 
metallicities were investigated: $-0.5$, 0, and $+0.5$~dex,
where [M/H]\,=\,0\,dex represents solar abundance and elements heavier than He
are scaled by $\pm 0.5$~dex in the other cases. A constant value of
2~km s$^{-1}$ as in the standard grid of Kurucz (1993b) was used for the
microturbulence. 

To transform the theoretical colours into the frame of observed colours
Kurucz (1993b) corrected the zero-point of his synthetic $uvby$ system so
as to match $c_1$, $m_1$, and $(b-y)$ of Vega.
A similar procedure is necessary to compare calculated $a$ values
from our synthetic $\Delta a$ system directly to observations. We have added $0.6$ to
the synthetic $a$ value computed from convolving the effective transmission
functions with the fluxes from ATLAS9 model atmospheres. The numerical
values we obtain are then close to Maitzen \& Vogt (1983, see their
Table 1 and Equation 1). We
obtain $a \sim 0.594$ for a main sequence star with $(b-y) = 0.000$ and solar
metallicity from our synthetic photometric system. For the $\Delta a$ system
itself the specific value of $a$ and thus any zero-point correction related
to it are irrelevant, because the quantity of interest is the difference of
a measured $a$ value to the normality line $a_0$. Hence, it is much more
important to show that our synthetic $\Delta a$ system recovers the observed
dependencies of the $\Delta a$ index from different indicators of effective
temperature, surface gravity, and metallicity.

Figure~\ref{Fig.colour_calib} illustrates the temperature (and gravity)
dependence of the $a$ index as a function of various experimental
indicators of \teff. For $(B-V)$ and $(g_1-y)$ the entire temperature range
is displayed while a cut-off was introduced for the other two cases
through requiring that $(b-y) \geqslant -0.02$ (i.e.\ $T_{\rm eff}
\lesssim 11000$~K) and $[$u-b$] \leqslant 1.38$ (i.e.\ $T_{\rm eff} \gtrsim
9250$~K). Models with $\log(g)=4$~dex have been connected with straight
lines to provide a proxy for the normality line $a_0$ of the
synthetic photometric system. Note that the output colours have been
rounded to 1~mmag accuracy as in Kurucz (1993b). The actual run of
the colours is continuous and smaller magnitude differences can hardly
be assigned a real physical meaning within the current state of modelling. 

For our comparison of the synthetic $\Delta a$ system to observational
ones we have used the results for Galactic field stars
(Maitzen \& Vogt 1983; Vogt et al.\ 1998). More than 1140 bright normal,
peculiar and related stars have been observed within four different
$\Delta a$ systems. These four systems are mainly distinguished by
slightly different filter transmission curves (Fig.~3 in Maitzen \& Vogt
1983). The differences for the $\Delta a$ values were found
to be in the range of 2 to 3~mmag. They find 
the following relation for the normality linegg:
$$ a_0 = G_0 + G_1(b-y) + G_2 (b-y)^2$$
with $G_{0} = 0.594$, and where $0.086 < G_{1} < 0.105$ as well as
$-0.050 < G_{2} < -0.150$ hold for $-0.120 < (b-y) < +0.200$. The 1$\sigma$
level around the normality line was found to be between 2.9 and 5.1~mmag.
These values are a superimposition of the internal measurement errors and
the (observed) natural bandwidth. On the other hand, the colours from model
atmospheres ranging the main sequence band from $\log(g)$ of [3.5,4.5] dex
for all models with a \teff\ of [7000,15000]~K and solar metallicity, and for
which $(b-y)\geq-0.02$, yield a $G_{0} = 0.591$ (with less than 0.5~mmag error),
while $G_{1} = 0.0985 \pm 0.0056$ and $G_{2} = -0.044 \pm 0.030$,
when fitting a least square parabola through the model colours (see
Fig.~\ref{Fig.colour_calib}). Hence, the $(b-y)$ dependence of the
experimental $\Delta a$ systems investigated in Maitzen \& Vogt (1983)
is reproduced very well.

For the [$u-b$] relation, Maitzen (1985) lists $G_{1}$\,=\,0.024 for 22 bright
unreddened stars with a 1$\sigma$ level of 4.5~mmag. For this correlation, the
results in Fig.~\ref{Fig.colour_calib} imply a more flat dependence of
$G_{1} = 0.0098 \pm 0.0018$ (and a $G_{2}$ of $-0.0022 \pm 0.00097$). However,
the rather small slope is very sensitive to the precise definition of the
sample: including giants with $\log(g)\geqslant 2.5$~dex would raise $G_{1}$
to 0.0161, whereas reducing the range of [$u-b$] from an upper limit of 1.38 to
1.20 while keeping only the main sequence band models with $\log(g) \geqslant
3.5$~dex, as in the first case, would increase it to $G_{1} = 0.0211 \pm
0.0028$. Thus, within the overall uncertainties expected for such kind of
a weak dependence on [$u-b$] the latter is reproduced sufficiently well.

Figure~\ref{Fig.colour_calib} also shows the correlation of $a$ with
the temperature indicators $(B-V)$ and $(g_1-y)$. Their dependency can easily
be studied as before, but is unlikely to reveal more information
beyond the uncertainties introduced by the colour transformation required
to compare the different filter systems used in observations and the
synthetic systems of Kurucz (1993b).

\subsection{Luminosity effects}    \label{Sect_lumin}

\begin{figure}
\plotone{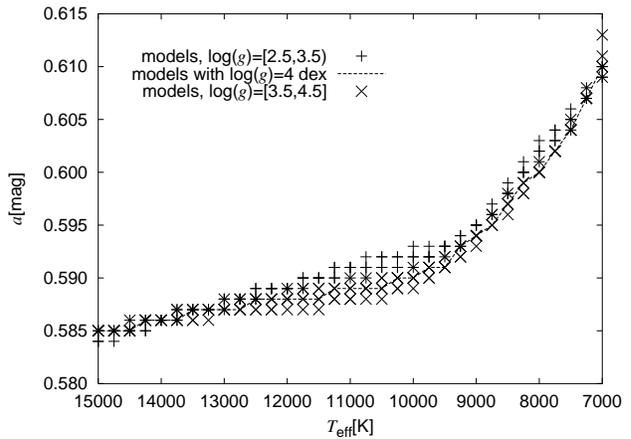}
\caption{Luminosity dependence of $a$ index. Colours from ATLAS9 type model
         atmospheres with \teff=[7000,15000]~K for $\log(g)$\,=\,[2.5,4.5]~dex and
         solar metallicity. Models above the main sequence band and thus at
         lower surface gravity are indicated by different symbols. Asterisks
         indicate the overlap of both ranges. Output accuracy is limited
         to 1~mmag (see text). Models with $\log(g)=4$~dex have been
         connected by straight lines to indicate the normality line $a_0$.}
\label{Fig.lumin_calib}
\end{figure}

Because each of the colour relations presented in Fig.~\ref{Fig.colour_calib}
is also affected by surface gravity within the range of effective
temperatures populated by the CP stars, we have looked at the direct
dependence of $a$ on \teff\ as well (see Fig.~\ref{Fig.lumin_calib}).
Clearly, models with lower surface gravity have higher $a$ values. The
effect of surface gravity on the $a$ index is largest for the late B stars
with effective temperatures around \teff\,$\sim$\,10500~K. The width of
the band of standard stars as induced by surface gravity for a given
metallicity is between 2 and 5~mmag. This confirms the results of
the previous subsection and is in agreement with the observational
data quoted therein.

We note here that the step size in $\log(g)$ for model sequences shown 
in both Fig.~\ref{Fig.lumin_calib} 
and \ref{Fig.metal_calib} is 0.25. However, as the output of the photometric indices 
has been truncated to 1~mmag, it turns out that the $\Delta a$ 
dependence on $\log(g)$ is too weak to show up more prominently. Hence, 
many models overlap in the Figures due the assumed output accuracy. The 
sensitivity of $\Delta a$ to $\log(g)$ slightly depends on \teff\ and 
metallicity, as the number of apparent points in Figs.~\ref{Fig.lumin_calib} 
and \ref{Fig.metal_calib} reveals as well.

\subsection{Metallicity effects}

\begin{figure}
\plotone{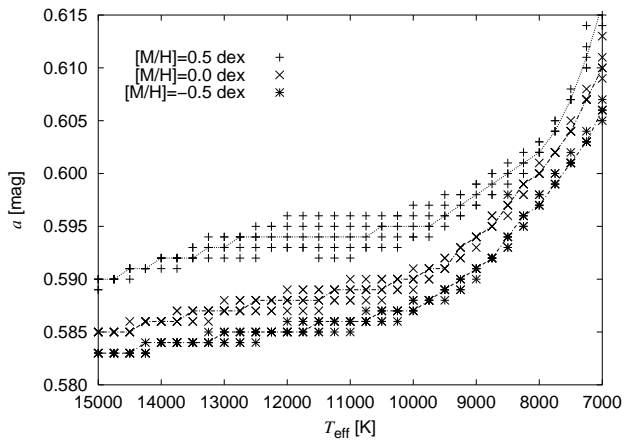}
\caption{Metallicity dependence of $a$ index. Colours from ATLAS9 type model
         atmospheres with \teff=[7000,15000]~K for $\log(g)$\,=\,[3.5,4.5]~dex.
         Metallicities are scaled with respect to the solar abundances
         used in Kurucz (1993b). Again, models with $\log(g)=4$~dex have been
         connected by straight lines to indicate the normality line $a_0$.}
\label{Fig.metal_calib}
\end{figure}

Metallicity has an effect on the $a$ index which is actually more important
than that of luminosity (surface gravity). Fig.~\ref{Fig.metal_calib}
compares the main sequence band for solar metallicity with models having
over- and underabundances of $\pm 0.5$~dex for all elements heavier than
He. We conclude that an underabundance of $-0.5$~dex as in the Magellanic
Clouds (Dirsch et al.\ 2000) yields a shift of the normality line of
$-3$~mmag. Notice that Maitzen, Paunzen \& Pintado (2001) found the
first extragalactic CP stars in the Magellanic Cloud using the $\Delta a$
system. The size of this shift is quite constant over the entire
\teff\ range relevant for CP stars and also within the entire luminosity
range expected for the main sequence band. On the other hand, an
overabundance of $+0.5$~dex yields a larger shift of between $+3$ and
$+6$~mmag with a maximum for the late B stars. This behaviour gives already
some hint on the nature of the flux depression at 5200~{\AA} in agreement
with Adelman et al.\ (1995) and Adelman \& Rayle (2000) who found that line
opacities of ATLAS9 models with metal overabundances of $+0.5$ and $+1.0$~dex
predict some extra line blanketing in this region. In turn, due to the good
agreement of ATLAS9 model fluxes in this wavelength region with observations
for mildly peculiar stars, which have underabundances or overabundances of 
up to about 0.5~dex (cf.\ Castelli \& Kurucz 1994; Adelman \& Rayle 2000),
and due to the satisfactory agreement of our synthetic $\Delta a$ system
with systems used in observations (see previous subsections), we can draw 
an important conclusion: application of $\Delta a$ photometry to the
Magellanic Clouds will lead only to a small bias, with a size of about
$-3$~mmag relative to the observations made for the solar neighbourhood, and
the same will hold for more remote targets that have a similar metallicity
range.

\section{Conclusions and outlook}  \label{Sect3}

In this first paper of our series we have
established a synthetic photometric $\Delta a$ system and confirm the observed dependency 
of the $a$ index as a function of various colour indices sensitive to 
the effective temperature and surface gravity variations within the Str\"omgren
$uvby\beta$ and Johnson $UBV$ photometric systems. 
Several
calibration relations are presented to confirm
that the new synthetic $\Delta a$ system provides a normality line and
features an average scatter along the main sequence for normal type stars
which is very close to the observed relations, if fluxes from ATLAS9
model atmospheres are used as input data for the synthetic
photometric system.
The metallicity dependence of the normality line of the $\Delta a$ system 
was computed for several grids of model
atmospheres for which the abundances of elements heavier than He had been scaled
by $\pm$0.5\,dex in order
to test for the effects of over- and underabundances. We estimate a lowering of $\Delta a$
by $\sim -3$~mmag assuming an average metallicity of $[{\rm Fe}/{\rm H}]=-0.5$~dex
compared to the Sun.
This is a typical value as found for Magellanic Clouds for which
the first CP stars have already been detected using the $\Delta a$ system. 
Thus, $\Delta a$ photometry is a viable tool to identify 
CP stars in samples with metallicities slightly different from the solar ones and 
it is well suited to draw statistically
meaningful conclusions about their distribution. It is hence a recommendable method
to find CP stars in other galaxies, too.

We intend to publish two follow-up papers. In these papers we will present
model atmospheres computed with individual abundances for a representative
sample of CP as well as $\lambda$ Bootis stars. For
these objects we will either confirm or redetermine the
input parameters (effective temperature, surface gravity, overall
metallicity and microturbulence) found
in the literature through comparisons with photometric, spectrophotometric,
and high
resolution spectra ($R$\,$\approx$\,20000) spectroscopic data. 
The final models obtained from this procedure will be
used to compute synthetic $\Delta a$ values which will be compared with 
individual photoelectric observations. 
The observed behaviour of $\Delta a$ will be shown to be very well reproduced for several
types of CP stars. Furthermore, a detailed statistical
analysis of the relative abundances for each star will be given.
This will illustrate which species contribute in the
different filters.

The first follow-up paper (paper {\sc ii}) will discuss the models
for Am and cool CP2 stars with effective temperatures below about
10000~K. The second one
will deal with a discussion of hotter CP stars
with spectral types earlier than A0 (paper {\sc iii}).
The atmospheres of these objects are different from those of A
stars. Among others, convection cannot play an essential
role any more and the effects of stratification become even more important than for
cooler objects.

\section*{Acknowledgments}

We thank our referee, Dr.~F.~Castelli for helpful comments and improvements.
This research was performed within the projects {\sl P13936-TEC} and 
{\sl P14984} of the Austrian Fonds zur F\"orderung der wissen\-schaft\-lichen
Forschung (FwF). We acknowledge use of the SIMBAD and GCPD astronomical data base.
For support during the
final stage of writing this paper FK acknowledges support
through the UK PPARC under grant PPA/G/O/1998/00576.

\label{lastpage}

\end{document}